\documentclass[floatfix,aps,prl,showpacs,amsmath,nofootinbib,
preprintnumbers,twocolumn]
{revtex4}

\usepackage{graphicx}
\usepackage{colordvi}
\begin{document}
\def\H{\,{\mathcal  H}}
\def\bea{\begin{eqnarray}}
\def\eea{\end{eqnarray}}

\title{Gauge-Invariant Temperature Anisotropies  and 
Primordial Non-Gaussianity}
\author{Nicola Bartolo $^{(1)}$, Sabino
Matarrese $^{(2,3)}$ and Antonio Riotto$^{(3)}$}
\address{$^{(1)}${\it Astronomy Centre, University of Sussex, 
Falmer, Brighton, BN1 9QH, U.K.}}
\address{$^{(2)}${\it Dipartimento di Fisica `Galileo Galilei', Universit\`a 
di Padova, 
via Marzolo 8, I-35131, Padova, Italy}}
\address{$^{(3)}${\it INFN, Sezione di Padova,
via Marzolo 8, I-35131, Padova, Italy}}

\date{\today}

\begin{abstract}
\noindent
We provide the gauge-invariant expression for large-scale 
cosmic microwave background
temperature fluctuations at second-order in perturbation theory. 
It enables to unambiguously define the nonlinearity parameter
$f_{\rm NL}$ which is used by experimental
collaborations to pin down the level of Non-Gaussianity 
in the temperature fluctuations. Furthermore, it contains 
a {\it primordial} term encoding all the information
about the Non-Gaussianity generated at primordial epochs and about 
the mechanism which gave rise to cosmological perturbations, thus 
neatly disentangling the primordial contribution to Non-Gaussianity
from the one caused by the post-inflationary evolution. 
\end{abstract}


\maketitle

\noindent
Inflation has become the dominant paradigm to understand the 
initial conditions for the density perturbations in the early 
Universe which are the seeds for the Large-Scale Structure (LSS)
and for the Cosmic Microwave Background (CMB) 
temperature anisotropies~\cite{lrreview}. 
In the inflationary picture, 
primordial density and gravity-wave fluctuations are
created from quantum fluctuations ``redshifted'' out of the horizon during an
early period of superluminal Universe expansion. 
Despite the simplicity of the inflationary paradigm, the mechanism
by which cosmological curvature (adiabatic) perturbations are generated is not
yet established. In the standard slow-roll inflationary scenario associated 
to one-single field, the inflaton, density 
perturbations are due to fluctuations of the inflaton itself when it
slowly rolls down along its potential. 
In the curvaton 
mechanism~\cite{curvaton} the final curvature perturbation $\zeta$ is 
produced from an initial isocurvature mode associated with the
quantum fluctuations of a light scalar (other than the inflaton), 
the curvaton, whose energy density is negligible during inflation. 
Recently, other mechanisms for the generation of cosmological
perturbations have been proposed, the inhomogeneous reheating scenario 
\cite{gamma1}, ghost-inflation 
\cite{ghost} and the D-cceleration scenario~\cite{dacc}, to mention 
a few.
A precise measurement of the spectral index $n_\zeta$ 
of comoving curvature perturbations will provide a powerful constraint
to slow-roll inflation models and the standard scenario
for the generation of cosmological perturbations
which predicts $|n_\zeta-1|$ significantly below 
unity. However, alternative mechanisms generically also predict 
a value of $n_\zeta$ very close to unity. Thus, even a
precise measurement of the spectral index will not 
allow us to efficiently discriminate among them.
On the other hand, the lack gravity-wave signals in CMB anisotropies will not
give us any information about the perturbation generation 
mechanism, since alternative mechanisms predict
an amplitude of gravity waves far too small to be
detectable by future experiments aimed at
observing the $B$-mode of the CMB polarization. 

There is, however, a third observable which will prove 
fundamental in providing information about the
mechanism chosen by Nature to produce the structures we see today. It
is the deviation from a Gaussian statistics, {\it i.e.}, the
presence of higher-order connected correlation functions of CMB anisotropies. 
Since for every scenario there exists a well defined prediction
for the strength of Non-Gaussianity (NG) and its shape as a function of the 
parameters, testing the NG level of primordial fluctuations is 
one of the most powerful probes of inflation
\cite{review} and is crucial to  
discriminate among different -- but otherwise indistinguishable - 
mechanisms. 
For instance, the single-field slow-roll inflation model 
itself produces negligible NG, 
and the dominant contribution comes from the evolution of 
the ubiquitous second-order perturbations after inflation, 
which is potentially detectable with future observations of CMB temperature 
and polarization anisotropies. This effect {\it must exist}
regardless of the inflationary models, setting the minimum NG level of 
cosmological perturbations. 
Therefore, if we do not find any evidence for this 
ubiquitous NG, then it will challenge our understanding of the 
evolution of cosmological perturbations at a deeper level.

Motivated by the extreme relevance of pursuing NG
in the CMB anisotropies, in this Letter we provide the exact 
expression for large-scale CMB temperature fluctuations at second order 
in perturbation theory. 
This expression has various virtues. First, it is gauge-invariant. Second, 
from it one can unambiguously extract the exact definition of the 
nonlinearity parameter $f_{\rm NL}$ which is used by the experimental
collaborations to pin down the level of NG 
in the temperature fluctuations. Third, it contains 
a ``primordial'' term encoding all the information
about the NG generated in primordial epochs, namely
during or immediately after inflation, and depends upon the various 
fluctuation generation mechanisms. As such, the expression 
neatly disentangles the primordial contribution to the NG
from that arising after inflation. Finally, 
the expression applies to all scenarios for the generation of cosmological
perturbations. 

In order to obtain our gauge-independent formula for the temperature 
anisotropies 
we first perturb a spatially flat Robertson-Walker background. Here we follow 
the formalism of Ref.~\cite{MMB} expanding 
metric perturbations  in a first and a second-order part as     
\begin{eqnarray} \label{metric1}
g_{00}&=&-a^2 \left( 1+2 \phi^{(1)}+\phi^{(2)} \right)\, ,
g_{0i}=a^2 \left( \hat{\omega}_i^{(1)}+\frac{1}{2} 
\hat{\omega}_i^{(2)} \right)\, ,
 \nonumber \\
g_{ij}&=&a^2\left[
(1 -2 \psi^{(1)} - \psi^{(2)})\delta_{ij}+
\left( \hat{\chi}^{(1)}_{ij}+\frac{1}{2}\hat{\chi}^{(2)}_{ij} \right)\right] 
\, ,
\end{eqnarray}
where the scale factor $a(\eta)$ is a function of the conformal time $\eta$.
The functions $\phi^{(r)}, \hat{\omega}_i^{(r)}, 
\psi^{(r)}$ and $\hat{\chi}^{(r)}_{ij}$, where $(r)=(1,2)$, stand for the 
$r$th-order perturbations of the metric. 
It is standard use to split the perturbations into the so-called scalar, 
vector and tensor parts,  
according to their transformation properties with respect to 
the $3$-dimensional space with metric $\delta_{ij}$, where scalar parts are 
related to a scalar potential, vector parts to transverse (divergence-free) 
vectors and tensor parts to transverse trace-free tensors. 
Thus $\phi$ and $\psi$ are scalar perturbations, 
and for instance, 
$\hat{\omega}_i^{(r)}=\partial_i\omega^{(r)}+\omega_i^{(r)}$, 
where $\omega^{(r)}$ is the scalar part and $\omega^{(r)}_i$ 
is a transverse vector, {\it i.e.} $\partial^i\omega^{(r)}_i=0$.
The metric perturbations will transform according to an infinitesimal 
change of coordinates. From now on we limit ourselves to a second-order time 
shift 
$\eta\rightarrow 
\eta-\alpha_{(1)}+\frac{1}{2} ({\alpha'_{(1)}}\alpha_{(1)}
-\alpha_{(2)})$, where a prime denotes differentiation w.r.t. conformal 
time.
In general a gauge corresponds to a choice of coordinates 
defining a slicing of spacetime into hypersurfaces (at fixed time $\eta$) 
and a threading into lines (corresponding to fixed spatial coordinates 
${\bf x}$), but in this Letter only the former is relevant so that
gauge-invariant can be taken to mean independent of the slicing~\cite{mw}.
For example, under the time shift, 
the first-order spatial curvature perturbation 
$\psi^{(1)}$ transforms as $
\psi^{(1)}\rightarrow\psi^{(1)}-{\mathcal H} \,\alpha_{(1)}$ 
(here ${\mathcal H}=a'/a$), 
while $\phi^{(1)} \rightarrow \phi^{(1)}+\alpha'_{(1)}
+{\mathcal H} \alpha^{(1)}$, 
$\hat{\omega}_i^{(1)} \rightarrow
\hat{\omega}_i^{(1)}-\partial_i \alpha^{(1)}$, and the traceless 
part of the spatial metric $\hat{\chi}^{(1)}_{ij}$ turns out 
to be gauge-invariant.
At second order in the perturbations 
we just give some useful examples like the 
transformation of the energy density and the curvature perturbation~\cite{MMB} 
$\delta^{(2)} \rho\rightarrow\delta^{(2)} \rho +\rho^\prime\alpha_{(2)} +
\alpha_{(1)}\left(\rho^{\prime\prime}\alpha_{(1)}
+\rho^\prime\alpha_{(1)}^\prime+2\delta^{(1)} \rho^\prime\right)$ and 
$\psi^{(2)}\rightarrow\psi^{(2)}
+2\alpha_{(1)}
\left(\psi^{(1)\prime}+
2{\mathcal H}\psi^{(1)}\right)
-\left({\mathcal H}'+2{\mathcal H}^2\right)\alpha^2_{(1)}
-{\mathcal H} \alpha_{(1)} \alpha_{(1)}^{\prime}
- \frac{1}{3}\left(2\hat{\omega}^{i}_{(1)}-
\alpha^{,i}_{(1)}\right)\alpha^{(1)}_{,i} -{\mathcal H} \alpha_{(2)}\,$ .
In particular, there exists an extension at second order of 
the well-known gauge-invariant variable 
$\zeta^{(1)}=-\psi^{(1)}-{\mathcal H} \frac{\delta^{(1)} \rho}{\rho'}$ 
(the curvature perturbation on uniform density hypersurfaces). 
It is given by $\zeta=\zeta^{(1)}+(1/2)\zeta^{(2)}$, where~\cite{lw,mw}
\begin{eqnarray}
\label{qqq}
&-&\zeta^{(2)}= 
\psi^{(2)}+{\mathcal H}\frac{\delta^{(2)}\rho}{\rho^\prime}
-2{\mathcal H}\frac{\delta^{(1)}\rho^\prime}{\rho^\prime}
\frac{\delta^{(1)}\rho}{\rho^\prime} 
-2\frac{\delta^{(1)}\rho}{\rho^\prime}{\psi}^{(1)\prime}
\nonumber \\
&-&4 {\mathcal H} \frac{\delta^{(1)}\rho}{\rho^\prime}{\psi}^{(1)}
+\left(\frac{\delta^{(1)} \rho}{\rho'}\right)^2 
\left({\mathcal H} \frac{\rho^{\prime\prime}}{\rho^\prime}-
{\mathcal H}^\prime-2{\mathcal H}^2\right) \, .
\end{eqnarray}
The key point here is that 
the  gauge-invariant comoving curvature perturbation
$\zeta^{(2)}$ 
remains  {\it constant} on superhorizon scales after it has been generated
and possible isocurvature perturbations are no longer present.
Therefore,  $\zeta^{(2)}$ 
provides all the necessary information about the
``primordial'' level of NG generated either during inflation, as in the 
standard scenario, or immediately after it, as in the curvaton scenario. 
Different scenarios
are characterized by different values of  $\zeta^{(2)}$, while
the post-inflationary nonlinear evolution due to gravity is 
common to all of them~\cite{BMR2,BMR3,BMR4,review}.
For example, 
in standard single-field inflation $\zeta^{(2)}$ is generated during inflation
and its value is $\zeta^{(2)}=2\left( 
\zeta^{(1)} \right)^2+{\cal O}\left(n_\zeta-1\right) $~\cite{ABMR,BMR2}. 

We now construct in a gauge-invariant way temperature anisotropies 
at second order. Temperature anisotropies beyond the linear regime   
have been calculated in Refs.~\cite{T2nd},  
following the photons path from last-scattering 
to the observer in terms of perturbed geodesics. 
The linear temperature anisotropies read~\cite{T2nd}
\begin{equation} 
\label{T1}
\frac{\Delta T^{(1)}}{T}=\phi^{(1)}_{\mathcal E} -v^{(1)i}_{\mathcal E}e_i
+ \tau^{(1)}_{\mathcal E}
-\int_{\lambda_{\mathcal O}}^{\lambda_{\mathcal E}} d\lambda A^{(1) \prime} \;,
\end{equation} 
where $A^{(1)}\equiv \psi^{(1)}+\phi^{(1)}+\hat{\omega}^{(1)}_i e^i-
\frac{1}{2}\hat{\chi}^{(1)}_{ij}e^i e^j$, the subscript 
${\mathcal E}$ indicates that quantities are evaluated at 
last-scattering, $e^i$ is a spatial unit vector 
specifying the direction of observation and the integral 
is evaluated along the line-of-sight parametrized by the affine parameter 
$\lambda$. Eq.~(\ref{T1}) includes the intrinsic fractional 
temperature fluctuation at emission  
$\tau_{\mathcal E}$, the Doppler effect due to emitter's velocity 
$v^{(1)i}_{\mathcal E}$ and the
gravitational redshift of photons, including the Integrated 
Sachs-Wolfe (ISW) effect. We omitted monopoles 
due to the observer ${\mathcal O}$ ({\it e.g.} the gravitational 
potential $\psi^{(1)}_{\mathcal O}$ evaluated at the event of observation), 
which, being independent of the angular coordinate, can be always recast  
into the definition of temperature anisotropies~\cite{HwangNoh}. 
Notice however that the physical meaning of 
each contribution in Eq.~(\ref{T1}) is not gauge-invariant, as 
the different terms are gauge-dependent. However, 
it is easy to show that the whole expression~(\ref{T1}) is 
gauge-invariant. 
Since the temperature $T$ is a scalar,     
the intrinsic temperature fluctuation transforms as 
$\tau_{\mathcal E}^{(1)}\rightarrow\tau_{\mathcal E}^{(1)}
+(T'/T)\alpha_{(1)}=\tau_{\mathcal E}^{(1)}
-{\mathcal H}\alpha_{(1)}$, 
having used the fact that the temperature scales as 
$T \propto a^{-1}$. Notice, instead, 
that the velocity $v^{(1)i}_{\mathcal E}$ 
does not change. Therefore, using the transformations of metric 
perturbations we find  
\begin{eqnarray}
\label{proof2}
\frac{\Delta T^{(1)}}{T} &\rightarrow &\frac{\Delta T^{(1)}}{T}+\alpha'_{(1)}
-\int_{\eta_{\mathcal O}}^{\eta_{\mathcal E}} 
d\eta \frac{d \alpha'_{(1)}}{d \eta}
=\frac{\Delta T^{(1)}}{T} + {\mathcal O}\, , \nonumber \\
&& 
\end{eqnarray}
where we have used the fact that the integral is evaluated along 
the line-of-sight which can be parametrized by the background 
geodesics $x^{(0) \mu}=
\left( \lambda, (\lambda_{\mathcal O}-\lambda_{\mathcal E}) e^i \right)$ 
(with $d \lambda/d \eta=1$), and the decomposition for the total 
derivative along the path 
for a generic function $f(\lambda,x^i(\lambda))$, 
$f'=\frac{\partial f}{\partial \lambda}=
\frac{d f}{d \lambda} + \partial_i f e^i$. 
Eq.~(\ref{proof2}) shows that the expression~(\ref{T1}) for 
first-order temperature 
anisotropies is indeed gauge-invariant 
(up to monopole terms related to the observer ${\mathcal O}$). 
Temperature anisotropies can be easily written in 
terms of particular combinations of perturbations which are manifestly 
gauge-invariant. For the gravitational potentials we consider 
the gauge-invariant 
definitions $\psi^{(1)}_{\rm GI}=\psi^{(1)}-{\mathcal H} \omega^{(1)}$ and 
$\phi^{(1)}_{\rm GI}=\phi^{(1)}+{\mathcal H} \omega^{(1)}+\omega^{(1)'}$.
For the $(0-i)$ component of the metric and the traceless part of 
the spatial metric we define $\omega_i^{(1) \rm GI}=\omega_i^{(1)}$ and $
\hat{\chi}^{(1) \rm GI}_{ij}=\hat{\chi}^{(1)}_{ij}$.
For the matter variables we use a 
gauge-invariant intrinsic temperature fluctuation
$\tau^{(1)}_{\rm GI}=\tau^{(1)}- {\mathcal H}\omega^{(1)}$, 
while the velocity itself is gauge-invariant 
$v^{(1)i}_{\rm GI}=v^{(1)i}$ under time shifts.
Following the same steps 
leading to Eq.~(\ref{proof2}) one gets the linear 
temperature anisotropies in Eq.~(\ref{T1}) in terms of 
these gauge-invariant quantities
\begin{eqnarray}
\label{T1GI}
& &\frac{\Delta T^{(1)}_{\rm GI}}{T}=\phi^{(1)}_{\rm GI} -v^{(1)i}_{\rm GI}e_i
+ \tau^{(1)}_{\rm GI} -\int_{\lambda_{\mathcal O}}^{\lambda_{\mathcal E}} 
d \lambda\,  A^{(1) \prime}_{\rm GI}\, ,
\end{eqnarray}  
where $A^{(1)}_{\rm GI}=
\phi^{(1)}_{\rm GI}+\psi^{(1)}_{\rm GI}+\omega_i^{(1) \rm GI} 
e_i-\frac{1}{2}\hat{\chi}^{(1) \rm GI}_{ij} e^i 
e^j$ and we omitted the subscript ${\mathcal E}$. 
For the primordial fluctuations we are interested in the large-scale 
modes set by the curvature perturbation $\zeta^{(1)}$. Defining a 
gauge-invariant density perturbation $\delta^{(1)}\rho_{\rm GI}=
\delta^{(1)} \rho+\rho' \omega^{(1)}$, we write the curvature 
perturbation as $\zeta^{(1)}_{\rm GI}=-\psi^{(1)}_{\rm GI}-
{\mathcal H} (\delta^{(1)} \rho_{\rm GI}/\rho')$. 
Since for adiabatic perturbations in the radiation ($\gamma$) and 
matter ($m$) eras
$(1/4)(\delta^{(1)} \rho_\gamma/\rho_{\gamma}) = (1/3)
(\delta^{(1)} \rho_m/\rho_{m})$, one can write the 
intrinsic temperature fluctuation as 
$\tau^{(1)}=(1/4)(\delta^{(1)} \rho_\gamma/\rho_{\gamma})=
-{\mathcal H} (\delta^{(1)} \rho/\rho')$ and a gauge-invariant
definition is $\tau^{(1)}_{\rm GI}=-{\mathcal H} (\delta^{(1)} 
\rho_{\rm GI}/\rho')$. 
In the large-scale limit, from Einstein equations, in the 
matter era $\phi^{(1)}_{\rm GI}=\psi^{(1)}_{\rm GI}=-\frac{3}{5}
\zeta^{(1)}_{\rm GI}$. 
Thus we obtain the large-scale limit of temperature anisotropies~(\ref{T1GI}) 
$\frac{\Delta T^{(1)}_{\rm GI}}{T}=
2\psi^{(1)}_{\rm GI} +\zeta^{(1)}_{\rm GI}= 
\psi^{(1)}_{\rm GI}/3$, i.e. the usual Sachs-Wolfe effect. 

At second order, the procedure is similar to the one described 
so long, though more lengthy and cumbersome. 
We only provide the reader with the  
main steps to get the final expression. The 
second-order temperature fluctuations 
in terms of metric perturbations read~\cite{T2nd}
\begin{eqnarray}
\label{T2}
& &\frac{\Delta T^{(2)}}{T}=\frac{1}{2}\phi^{(2)}_{\mathcal{E}}
-\frac{1}{2}\left( \phi_{\mathcal{E}}^{(1)} \right)^2
-\frac{1}{2}v^{(2)i}_{\mathcal E}e_i
+ \frac{1}{2}\tau^{(2)}_{\mathcal E}
-I_2(\lambda_{\mathcal{E}}) \nonumber\\
&+&\left(I_1(\lambda_{\mathcal{E}})+v^{(1)i}_{\mathcal{E}} e_i\right)
\left(-\phi^{(1)}_{\mathcal{E}}-\tau^{(1)}_{\mathcal E}
+v^{(1)i}_{\mathcal{E}} e_i
+I_1(\lambda_{\mathcal{E}})\right)\nonumber\\
&+&x^{(1)0}_{\mathcal{E}} A ^{(1)'}_{\mathcal{E}}+(x^{(1)j}_{\mathcal{E}}
+x^{(1)0}_{\mathcal{E}} e^j)\left(\phi^{(1)}_{,j}-v^{(1)}_{i,j} e^i+
\tau^{(1)}_{,j}\right)_{\mathcal{E}} \nonumber\\
&-&\frac{1}{2}v^{(1)}_{{\mathcal{E}}i} v^{(1)i}_{\mathcal{E}}
+\phi^{(1)}_{\mathcal{E}}\tau^{(1)}_{\mathcal E}+
\frac{\partial \tau^{(1)}}{\partial d^i}d^{(1)i} 
-v^{(1)i}_{\mathcal{E}} e_i \phi^{(1)}_{\mathcal{E}}
\nonumber\\
&+&v^{(1)}_{{\mathcal{E}}i}\left(-\hat{\omega}^{(1)i}_{\mathcal{E}}
- I_1^i(\lambda_{\mathcal {E}})\right)\, . 
\end{eqnarray}
Here $I_2$ is the second-order ISW~\cite{T2nd} 
$I_2(\lambda_{\mathcal{E}})=\int_{\lambda_{\mathcal{O}}}
^{\lambda_{\mathcal{E}}} d\lambda [\frac{1}{2}A^{(2)'}
-(\hat{\omega}^{(1)'}_i-\hat{\chi}^{(1)'}_{ij} e^j)
(k^{(1)i}+e^i k^{(1)0})+2 k^{(1)0} A^{(1)'} +2 \psi^{(1)'} A^{(1)}
+x^{(1)0} A^{(1)''}+x^{(1)i} A^{(1)'}_{,i}]$, 
where $A^{(2)}\equiv \psi^{(2)}+\phi^{(2)}+\hat{\omega}^{(2)}_i e^i-
\frac{1}{2}\hat{\chi}^{(2)}_{ij}e^i e^j$,
while $k^{(1)0}(\lambda)=-2 \phi^{(1)}
-\hat{\omega}^{(1)i} e_i +I_1(\lambda)$ and 
$k^{(1)i}(\lambda)=-2 \phi^{(1)} e^i-
\hat{\omega}^{(1)i}+\hat{\chi}^{(1)ij} e_j
- I_1^i(\lambda)$ are the photon wave vectors, with 
$I_1(\lambda)$ given by the integral in Eq.~(\ref{T1}) and $I_1^i(\lambda)$ 
is obtained from the same integral replacing the time derivative with a
spatial gradient. 
Finally in Eq.~(\ref{T2}) 
$x^{(1)0}(\lambda)=
\int_{\lambda_{\mathcal{O}}}^{\lambda}d\lambda'
\left[-2 \phi^{(1)}-\hat{\omega}^{(1)}_i e^i+(\lambda-\lambda')
A^{(1)'}\right]$ and 
$x^{(1)i}(\lambda)=
-\int_{\lambda_{\mathcal{O}}}^{\lambda}d\lambda'
\left[2 \psi^{(1)} e^i+\hat{\omega}^{(1)i}-\hat{\chi}^{(1)ij} e_j
+(\lambda-\lambda')A^{(1),i}\right]$ are the geodesics at first 
order, and $d^{(1)i}=e^i-\frac{e^i-k^{(1)i}}{|e^i-k^{(1)i}|}$ is the 
direction of the photon emission. As usual we have omitted the monopole 
terms due to the observer. Using the transformation rules of 
Ref.~\cite{MMB}, it is possible to check that the expression~(\ref{T2}) 
is gauge-invariant. We can express the 
second-order anisotropies in terms of explicitly gauge-invariant 
quantities, whose definition proceeds as for the linear case, 
by choosing the shifts $\alpha^{(r)}$ such that 
$\omega^{(r)}=0$. For example, we consider the gauge-invariant gravitational 
potential~\cite{BMR4} 
\begin{eqnarray}
\label{PhiGI}
&\phi^{(2)}_{\rm GI}&=\phi^{(2)}+\omega^{(1)}\left[2\left(
\psi^{(1)'}+2\frac{a'}{a}\psi^{(1)}\right)+\omega^{(1) \prime \prime} 
+ 5 \frac{a'}{a}\omega^{(1) '}\right.\nonumber \\
&+&\left. \left( {\mathcal H}'+2 {\mathcal H}^2 \right)
\omega^{(1)}\right]
+2\omega^{(1)'}\left(2\psi^{(1)}+\omega^{(1)'}\right)+
\frac{1}{a} \left( a\alpha^{(2)} \right)' \, , \nonumber \\
\end{eqnarray}
where $\alpha^{(2)}=\omega^{(2)}+\omega^{(1)}\omega^{(1)'}
+\nabla^{-2}\partial^i[-4\psi^{(1)}\partial_i\omega^{(1)}-2
\omega^{(1)'}\partial_i\omega^{(1)}]$. 
Expressing the second-order temperature anisotropies~(\ref{T2}) 
in terms of our gauge-invariant quantities and taking the large-scale limit 
we find
$\Delta T^{(2)}_{\rm GI}/T=(1/2)\phi^{(2)}_{\rm GI}
-(1/2)\left( \phi_{\rm GI}^{(1)} \right)^2
+ (1/2)\tau^{(2)}_{\rm GI}+\phi^{(1)}_{\rm GI} \tau^{(1)}_{\rm GI}$
(having dropped the subscript ${\mathcal E}$), and the 
gauge-invariant intrinsic temperature fluctuation at emission is 
$\tau^{(2)}_{\rm GI}=(1/4) 
(\delta^{(2)} \rho^{\rm GI}_\gamma/\rho_\gamma)
-3( \tau^{(1)}_{\rm GI})^2$. 
We have dropped those terms which represent integrated 
contributions and other second-order 
small-scale effects that can be distinguished from the large-scale part 
through their peculiar scale dependence.
At this point we make use of Einstein's equations. We take the  
expression for $\zeta^{(2)}$ in Eq.~(\ref{qqq}), and we use 
the $(0-0)$ component and the traceless part of the $(i-j)$ Einstein's 
equation at second 
order (see Eqs.~(153) and~(155) of 
Ref.~\cite{review}). Thus, on large scales we find that the temperature 
anisotropies are given by
\begin{equation}
\label{main}
\frac{\Delta T^{(2)}_{\rm GI}}{T}=
\frac{1}{18} \left( \phi^{(1)}_{\rm GI} \right)^2 
-\frac{{\mathcal K}}{10}-\frac{1}{10} \left[ \zeta^{(2)}_{\rm GI}-
2 \left( \zeta^{(1)}_{\rm GI} \right)^2 \right]\, , 
\end{equation} 
where we have defined a kernel  ${\mathcal K}=
10 \nabla^{-4} \partial_i \partial^j 
 (\partial^i \psi^{(1)} \partial_j
\psi^{(1)}) -\nabla^{-2} 
( \frac{10}{3} \partial^i \psi^{(1)} \partial_i \psi^{(1)} )$. 
Eq.~(\ref{main}) is the main result of this Letter. It clearly shows that 
there are two contributions to the final nonlinearity in the 
large-scale temperature anisotropies. 
The  contribution, $[\zeta^{(2)}_{\rm GI}-
2 ( \zeta^{(1)}_{\rm GI} )^2]$, 
comes from the ``primordial'' conditions set during or after inflation. 
They are encoded in 
the curvature perturbation $\zeta$ which remains constant 
once it has been generated. 
The remaining part of Eq.~(\ref{main}) describes the post-inflation
processing of the primordial non-Gaussian signal due to the nonlinear 
gravitational dynamics, including also second-order corrections at last 
scattering to the Sachs-Wolfe effect~\cite{T2nd}. 
Thus, the expression in 
Eq.~(\ref{main}) allows to neatly disentangle the 
primordial contribution to NG from that coming from that arising 
after inflation. While the nonlinear evolution after inflation is the same in 
each scenario, the primordial content will depend on
the particular mechanism generating the perturbations.
We parametrize the primordial NG in the terms of the conserved curvature 
perturbation (in the radiation or matter dominated epochs)
$\zeta^{(2)}=2a\left(\zeta^{(1)}\right)^2$, 
where $a$ depends on the physics of a given scenario. For example, in the 
curvaton case $a=(3/4r)-r/2$, where 
$r \approx (\rho_\sigma/\rho)_{\rm D}$ is the relative   
curvaton contribution to the total energy density at curvaton 
decay~\cite{review}. In the minimal picture for the inhomogeneous 
reheating scenario, $a=1/4$. For the other scenarios we refer 
the reader to~Ref.\cite{review}. 
From Eq.~(\ref{main}) we can extract the nonlinearity parameter 
$f_{\rm NL}$ which is usually adopted to phenomenologically 
parametrize the NG level of cosmological perturbations and 
has become the standard quantity to be observationally
constrained by CMB experiments~\cite{ks,k}.
The definition of $f_{\rm NL}$ adopted in the analyses performed in 
Refs.~\cite{ks,k} goes through the conventional Sachs-Wolfe formula 
$\Delta T/T= - \Phi/3$ 
where $\Phi$ is Bardeen's potential 
\cite{Bardeen80}, which is conventionally expanded as 
(up to a constant offset, which only affects the temperature monopole)
$\Phi = \Phi_{\rm L} + f_{\rm NL} * \left(\Phi_{\rm L}\right)^2$, 
with  $\Phi_{\rm L} = - \phi^{(1)}_{\rm GI}$. Here the $\star$ product 
reminds the fact that the nonlinearity parameter might have a non-trivial 
scale dependence~\cite{review}. 
Therefore, using  $\zeta^{(1)}=-\frac{5}{3} 
\psi^{(1)}_{\rm GI}$ during matter domination, 
from Eq.~(\ref{main}) we read the nonlinearity parameter in 
momentum space  
\begin{equation}
\label{f_NL}
f_{\rm NL}({\bf k}_1,{\bf k}_2)
=-\left[ \frac{5}{3} \left(1-a \right) 
+\frac{1}{6}-\frac{3}{10} {\mathcal K} 
\right]+1\, 
\end{equation}
where ${\mathcal K}=10\, ({\bf k}_1 \cdot {\bf k}_3) 
({\bf k}_2 \cdot {\bf k}_3)/k^4 -\frac{10}{3}
{\bf k}_1 \cdot {\bf k}_2/k^2$ 
with ${\bf k}_3+{\bf k}_1+{\bf k_2}=0$ and $k=\left|
{\bf k}_3\right|$. In fact the formula
~(\ref{f_NL}) already accounts for an additional nonlinear effect 
entering in the CMB angular $3$-point function from the angular 
averaging performed with a perturbed line element 
$d \Omega (1-2 \psi^{(1)}_{\rm GI})$ ~\cite{review}, implying a $+1$ shift 
in $f_{\rm NL}$. 
In particular within the standard scenario 
where cosmological perturbations are due to the inflaton the 
primordial contribution to NG is given by 
$a=1-\frac{1}{4} (n_{\zeta}-1)$~\cite{ABMR,BMR2}, where the 
spectral index is expressed in terms of the usual slow-roll parameters as
$n_{\zeta}-1=-6 \epsilon +2 \eta$~\cite{lrreview}. The 
nonlinearity parameter from inflation now reads
\begin{equation}
f^{\rm inf}_{\rm NL}=-\frac{5}{12} (n_{\zeta}-1)  
+\frac{5}{6}+\frac{3}{10} {\mathcal K}\, . 
\end{equation} 
Therefore the main NG contribution comes from the 
post-inflation evolution of the second-order perturbations 
which give rise to order-one coefficients, while the primordial 
contribution is proportional to $|n_{\zeta}-1|\ll 1$. 
This is true even in   
the ``squeezed'' limit first discussed by Maldacena~\cite{Maldacena}, 
where one of the wavenumbers is much smaller than the other two, 
\emph{e.g.} $k_1 \ll k_{2,3}$ and ${\cal K}\rightarrow 0$. 

{\it Acknowledgments.} We thank J.~Peebles for spurring our efforts in 
disentangling the primordial (inflationary) NG in CMB anisotropies.


\begin{thebibliography}{99}

\bibitem{lrreview} 
D.~H.~Lyth and A.~Riotto, Phys.\ Rept.\ 314 (1999) 1.


\bibitem{curvaton} K.~Enqvist and M.~S.~Sloth,
Nucl.\ Phys.\ B 626 (2002) 395; D.~Lyth and D.~Wands,
Phys.\ Lett.\ B 524 (2002) 5; T.~Moroi and T.~Takahashi, Phys.\ Lett.\ B 522 
(2001) 215 [Erratum-ibid. B 539 (2002) 303].

\bibitem{gamma1} G.~Dvali, A.~Gruzinov and M.~Zaldarriaga,
Phys.\ Rev.\ D 69 (2004) 023505.

\bibitem{ghost} N.~Arkani-Hamed, H.~C.~Cheng, M.~A.~Luty and S.~Mukohyama,
arXiv:hep-th/0312099.

\bibitem{dacc}
E.~Silverstein and D.~Tong,
arXiv:hep-th/0310221.

\bibitem{review} 
N.~Bartolo, E.~Komatsu, S.~Matarrese and A.~Riotto,
arXiv:astro-ph/0406398, to apperar in Phys. Rept.
\bibitem{MMB} 
S.~Matarrese, S.~Mollerach and M.~Bruni, Phys.\ Rev.\ D 58 (1998)
043504.

\bibitem{mw} 
K.~A.~Malik and D.~Wands, Class.\ Quant.\ Grav.\ 21 (2004) L65.

\bibitem{lw} 
D.~H.~Lyth and D.~Wands, Phys.\ Rev.\ D 68 (2003) 103515.

\bibitem{BMR2}
N.~Bartolo, S.~Matarrese and A.~Riotto, JHEP\ 0404 (2004) 006.

\bibitem{BMR3}
N.~Bartolo, S.~Matarrese and A.~Riotto, 
Phys.\ Rev.\ D 69 (2004) 043503.

\bibitem{BMR4}
N.~Bartolo, S.~Matarrese and A.~Riotto, 
JCAP\ 0401 (2004) 003.

\bibitem{ABMR} 
V.~Acquaviva, N.~Bartolo, S.~Matarrese and A.~Riotto,
Nucl.\ Phys.\ B {\bf 667} (2003) 119.


\bibitem{T2nd}
T.~Pyne and S.~M.~Carroll, Phys.\ Rev.\ D 53 (1996) 2920; 
S.~Mollerach and S.~Matarrese, Phys.\ Rev.\ D 56 (1997) 4494.

\bibitem{HwangNoh}
J.~Hwang and H.~Noh, Phys.\ Rev.\ D 59 (1999) 067302.

\bibitem{ks}
E.~Komatsu and D.~N.~Spergel,
Phys.\ Rev.\ D 63 (2001) 063002.

\bibitem{k}
E.~Komatsu, et al., Astrophys.\ J.\ Suppl.\  148 (2003) 119.

\bibitem{Bardeen80}
J.~M.~Bardeen, Phys.~Rev.~D~22, (1980) 1882. 


\bibitem{Maldacena}
J.~Maldacena, JHEP\ 0305 (2003) 013.


\end{thebibliography}
\end{document}